\documentclass{PoS}
\usepackage{graphicx}

\title{Cosmic rays and the magnetic field of the nearby starburst galaxy NGC\,253}

\ShortTitle{The radio halo of NGC\,253}

\author{Volker Heesen and Ralf-J\"urgen Dettmar\\
Ruhr-University Bochum, Germany\\
E-mail: \email{heesen@astro.rub.de},
	\email{dettmar@astro.rub.de}}
\author{Marita Krause and Rainer Beck\\
Max Planck Institute for Radio Astronomy, Germany\\
E-mail: \email{mkrause@mpifr-bonn.mpg.de},
	\email{rbeck@mpifr-bonn.mpg.de}}

\abstract{Using radio polarimetry we study the connection between the transport of cosmic rays (CR's), the three-dimensional magnetic field structure, and features of other ISM phases in the halo of NGC\,253. We present a new sensitive radio continuum map of NGC\,253 obtained from combined VLA and Effelsberg observations at $\lambda6.2\,{\rm cm}$. We find a prominent radio halo with a scaleheight of the thick radio disk of 1.7\,kpc. The linear dependence between the \emph{local} scaleheight of the vertical continuum emission and the cosmic ray electron (CRE) lifetime requires a vertical CR bulk speed of $270\,{\rm km\,s^{-1}}$.\\
The magnetic field structure of NGC\,253 resembles an ``X''-shaped configuration where the orientation of the large-scale magnetic field is plane-parallel only in the inner regions of the disk and at small distances from the galactic midplane. At larger galactocentric radii and further away from the midplane the vertical component becomes important. This is most clearly visible at the location of the ``radio spur'' southeast of the nucleus, where the magnetic field orientation is almost vertical. We made a simple model for the dominant toroidal $(r,\phi)$ magnetic field component using a spiral magnetic field with prescribed inclination and pitch angle. The residual poloidal $(r,\phi,z)$ magnetic field component which was revealed by subtracting the model from the observations shows a distinct ``X''-shaped magnetic field orientation centered on the nucleus. The orientation angle of the poloidal magnetic field is consistent with a magnetic field transport described by the superposition of the vertical CR bulk speed and the rotation velocity.\\
Hence, we propose a \emph{disk wind} which transports cosmic rays, magnetic field, and (partially) ionized gas from the disk into the halo.}

\FullConference{From planets to dark energy: the modern radio universe\\
		 October 1-5 2007\\
		 University of Manchester, Manchester, UK}

\begin{document}
\section{Observations}
We have conducted an extensive radio continuum multiwavelength study both with the VLA interferometer\footnote{The VLA (Very Large Array) is operated by the NRAO (National Radio Astronomy Observatory).} and the 100-m Effelsberg telescope\footnote{The Effelsberg 100-m telescope is operated by the Max-Planck Institut f\"ur Radioastronomie (MPIfR).}. A mosaic at $\lambda6.2\,{\rm cm}$ consisting of 15 pointings has been observed with the VLA.
\begin{figure}
\begin{minipage}[b]{0.6\textwidth}
\resizebox{0.95\hsize}{!}{\includegraphics{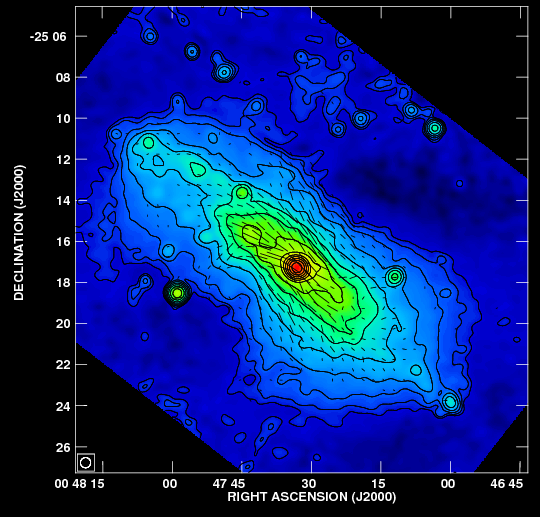}}
\end{minipage}
\hfill
\begin{minipage}[b]{0.4\textwidth}
\caption{Total power radio continuum obtained from the combined VLA + Effelsberg observations at $\lambda 6.2\,{\rm cm}$. The overlaid vectors indicate the intrinsic orientation of the resolved regular magnetic field. The length of the vectors is defined by 1\,arcsec to be equivalent to $12.5\,{\mu\rm Jy/beam}$ of polarized intensity. Contours are at 3, 6, 12, 24, 48, 96, 192, 384, 768, 1536, 3077, 6144, 12288, and 24576 $\times$ $30\,{\rm\mu Jy/beam}$. The size of the beam is 30\,arcsec (HPBW).}
\label{fig:n253cm6ve_tpaf_dss_b30}
\end{minipage}
\end{figure}
The strong nuclear continuum point source has been taken into account in our data reduction: we cleaned our single-dish maps in order to correct for sidelobes and applied a correction for the instrumental polarization. The instrumental polarization caused by the off-axis placement of the nuclear point source in the VLA primary beam has been addressed by a specially tailored polarization calibration. The missing zero-spacing flux of the VLA mosaic has been filled up with the $\lambda6.2\,{\rm cm}$ Effelsberg map.
\section{The cosmic ray distribution}
We find a prominent radio halo with a thin and a thick radio disk with a scaleheight of 0.3\,kpc and 1.7\,kpc at $\lambda6.2\,{\rm cm}$, respectively. The steepening of the spectral index with increasing distance from the disk suggests synchrotron radiation to be the dominant energy loss of the cosmic ray electrons (CRE's). The linear dependence between the \emph{local} scaleheight and the CRE lifetime requires a vertical CR bulk speed of $(270\pm30)\,{\rm km\,s^{-1}}$. This explains the dumbbell shape visible in the total power distribution presented in Fig.~\ref{fig:n253cm6ve_tpaf_dss_b30}, where the smallest scaleheights are found in the inner regions of the disk where the magnetic field strength is largest. The CR bulk speed is close to the escape velocity of $v_{\rm esc}=280\,{\rm km\,s^{-1}}$, where we used $v_{\rm esc}=\sqrt{2}\cdot v_{\rm rot}$ with a rotation velocity of $v_{\rm rot}=200\,{\rm km\,s^{-1}}$. The kinetic energy of the CR gas is thus possibly high enough to drive a galactic wind which eventually leaves the gravitational potential.
\section{The magnetic field structure}
\begin{figure}
\begin{minipage}[b]{0.6\textwidth}
\resizebox{0.95\hsize}{!}{\includegraphics{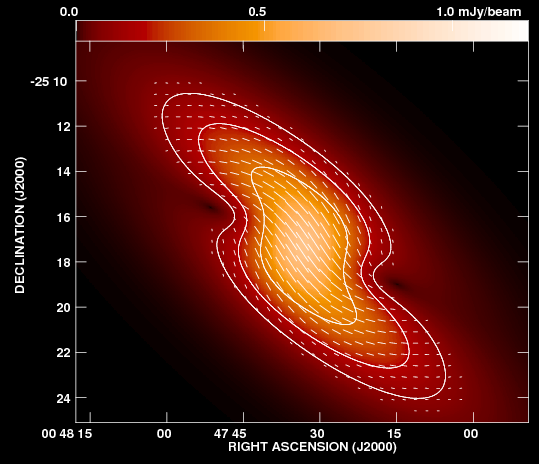}}
\end{minipage}
\hfill
\begin{minipage}[b]{0.4\textwidth}
\caption{Axisymmetric model for the toroidal magnetic field with a pitch angle of $25^\circ$ of the magnetic field spiral. Contours are at 3, 6, and 12 $\times$ $30\,{\mu\rm Jy/beam}$. The overlaid vectors indicate the intrinsic orientation of the resolved regular magnetic field. The length of the vectors is defined by 1\,arcsec to be equivalent to $12.5\,{\mu\rm Jy/beam}$ polarized intensity. The size of the beam is 30\,arcsec (HPBW).}
\label{fig:pa25_g700_pia}
\end{minipage}
\end{figure}

Using the revised equipartition formula of \cite{beck_05a} we derive a total magnetic field strength in the disk between $9\,{\rm\mu G}$ (outer part) and $16\,{\rm\mu G}$ (inner part). The distribution of the Rotation Measure (RM) shows the disk and halo to be split in a positive northeastern and negative southwestern part. From the comparison with the velocity field following the method described in \cite{krause_98a} we find the radial component of the spiral magnetic field to be directed inwards. Hence, we constructed a simple model for the large-scale magnetic field using an axisymmetric spiral magnetic field with a prescribed inclination and pitch angle (Fig.~\ref{fig:pa25_g700_pia}).
We find good agreement between the model and the observed magnetic field configuration. In order to get the distribution of the poloidal magnetic field we subtracted the model from the observed polarization. We find an ``X''-shaped pattern centered on the nucleus which correlates well with the extra-planar H$\alpha$ emission  and soft X-ray emission. This significant poloidal magnetic field component supports the transport of cosmic rays along the field lines connecting the disk with the halo. The orientation angle of the poloidal magnetic field can be explained by a \emph{disk wind} blowing from the disk into the halo as the superposition of the vertical CR bulk speed and the rotation velocity of the disk. This work will be fully presented in \cite{heesen_08a}, \cite{heesen_08b}, and \cite{heesen_08c}.

\end{document}